\documentclass[preprint,showpacs,preprintnumbers,amsmath,amssymb]{revtex4}

\usepackage{graphicx}
\usepackage{dcolumn}
\usepackage{bm}

\begin{document}

\title{Globally Anisotropic High Porosity Silica Aerogels}

\author{J. Pollanen}
\author{K. Shirer}
\author{S. Blinstein}
\author{J.P. Davis}
\author{H. Choi}
\author{T.M. Lippman}
\author{W.P. Halperin}
\affiliation{Department of Physics and Astronomy, Northwestern
University, Evanston,
IL 60208, USA}

\author{L.B. Lurio}
\affiliation{Department of Physics, Northern Illinois University,
DeKalb, IL 60115,
USA}

\date{\today}

\begin{abstract} 
We discuss two methods by which high porosity silica
aerogels can
be engineered to exhibit global anisotropy.  First, anisotropy can be
introduced
with axial strain.  In addition, intrinsic anisotropy can result
during growth
and drying stages and, suitably controlled, it can be correlated with preferential radial
shrinkage in
cylindrical samples.  We have performed small angle X-ray scattering (SAXS) to
characterize these two types of anisotropy.  We show that global anisotropy
originating from either strain or shrinkage leads to optical
birefringence and that
optical cross-polarization studies are a useful characterization of
the uniformity of the
imposed global anisotropy.
\end{abstract}


\maketitle

\section{INTRODUCTION} 
High porosity silica aerogel is a
transparent, low density, material consisting of aggregates of silica
particles, often described as strands, with a diameter of $\sim 30-50$ {\AA} and
average separation, or correlation length, $\xi \cong 100-1000$ {\AA}\cite{Fri88}.  
The microstructure of the aerogel is
inherently inhomogeneous on the scale of the correlation length, which
we take to be the
upper cut-off of a fractal distribution of the silica particles.
Fractal distributions
follow naturally from the physical process of diffusion limited
aggregation that
occurs during gel formation and have been characterized with small
angle scattering
experiments (visible light\cite{Hun84}, X-ray\cite{Sch86}, and
neutrons\cite{Vac88}) as well as having been
simulated numerically\cite{Has94}\cite{Wit81}.  In the present work we are
particularly
interested in long length scale inhomogeneity, specifically the
contolled introduction of
global anisotropy into silica aerogels and its characterization.  In
the following we
will use the term isotropic to refer to an aerogel having no
preferred direction on
average over macroscopic specimens that are typically of mm size or larger.

Silica aerogels with
porosities up to
$\sim 99 \%$ are of interest in a wide variety of scientific and
engineering disciplines
including high-energy physics, astrophysics, material science,
condensed matter physics
and chemical engineering.  Aerogels find application in particle
physics to detect
Cerenkov radiation\cite{Can74} and have been used by NASA to collect
particulate
matter from comet 81P/Wild 2\cite{Bak06}\cite{Bro06}.  Silica
aerogels have seen widespread application in materials science and
because of their
fractally correlated structure\cite{Sch86} have basic scientific
significance.  Aerogels
are used in the manufacture of ceramic materials\cite{Hru90}.  They
are important in the
study of critical phenomena and exploration of impurity effects in
quantum fluids
such as
$^4$He\cite{Cha96} and
$^3$He\cite{Hal04} as well as to understand the influence of
impurities on liquid crystal
phase transitions\cite{Cra96}\cite{Cla93}.  In addition, hybrid
materials based on silica aerogels can be engineered with metal, or
metal oxide,
particles incorporated into the aerogel framework for the purpose of producing
heterogeneous catalysts\cite{Mil94}.

Global anisotropy of the aerogel is an important consideration in many of these
systems.  In general, structural characterization is necessary to
understand the behavior
of fluids in porous media such as aerogels.
This is also true for anisotropic structures which can lead to
anisotropic diffusion
and flow and therefore be of consequence in many of the aforementioned applications.
For example, the ordered phases of liquid crystals in
aerogel are inherently anisotropic\cite{Cra96} and we can expect that
they will be
modified by global anisotropy\cite{Jac99}.
The physical and optical properties of aerogels are modified
by anisotropy, as we
discuss in this work, and this may be important in some applications
such as
transparent thermal insulation\cite{Emm95}.    More generally,
anisotropic aerogels are
members of a larger group of fascinating anisotropic porous media
including sedimentary
rock\cite{Win82}, bone\cite{Phe98}, and cementitious materials\cite{Min81}.

Our motivation for the present study is linked to our interest in a basic
physics problem concerning impurity effects on superfluid $^3$He.  It
is thought that
anisotropy will play a role in determining which phases of superfluid
$^3$He in aerogel
are energetically stable\cite{Aoy05,Dav06} and that anisotropy will
influence the
formation of textures of the superfluid order parameter.  Here we report that
high porosity silica
aerogels can be engineered to exhibit global anisotropy and we
discuss small angle X-ray
scattering (SAXS) and optical polarization methods for
characterization of this
anisotropy.

\section{SAMPLES} 
The aerogel samples
discussed in this
work were grown at Northwestern University via the ``one-step'' sol-gel
method\cite{Tei85}.  A silicon oxide precursor, tetramethyl
orthosilicate (TMOS), is
dissolved in methanol and hydrolyzed. Ammonia is used as a catalyst.
\begin{center} Si(CH${_2}$OH)${_4}$ $+$ 2H${_2}$O $\rightarrow$ SiO${_2}$ $+$
4CH${_3}$OH
\end{center}
This reaction results in a wet gel, known as alcogel.  The alcogel is then supercritically dried following a method that is based on the rapid supercritical extraction process (RSCE) of J.F. Poco {\em et al.}\cite{Poc96}.

\begin{figure}[h]
\centerline{\includegraphics[width=0.5\textwidth]{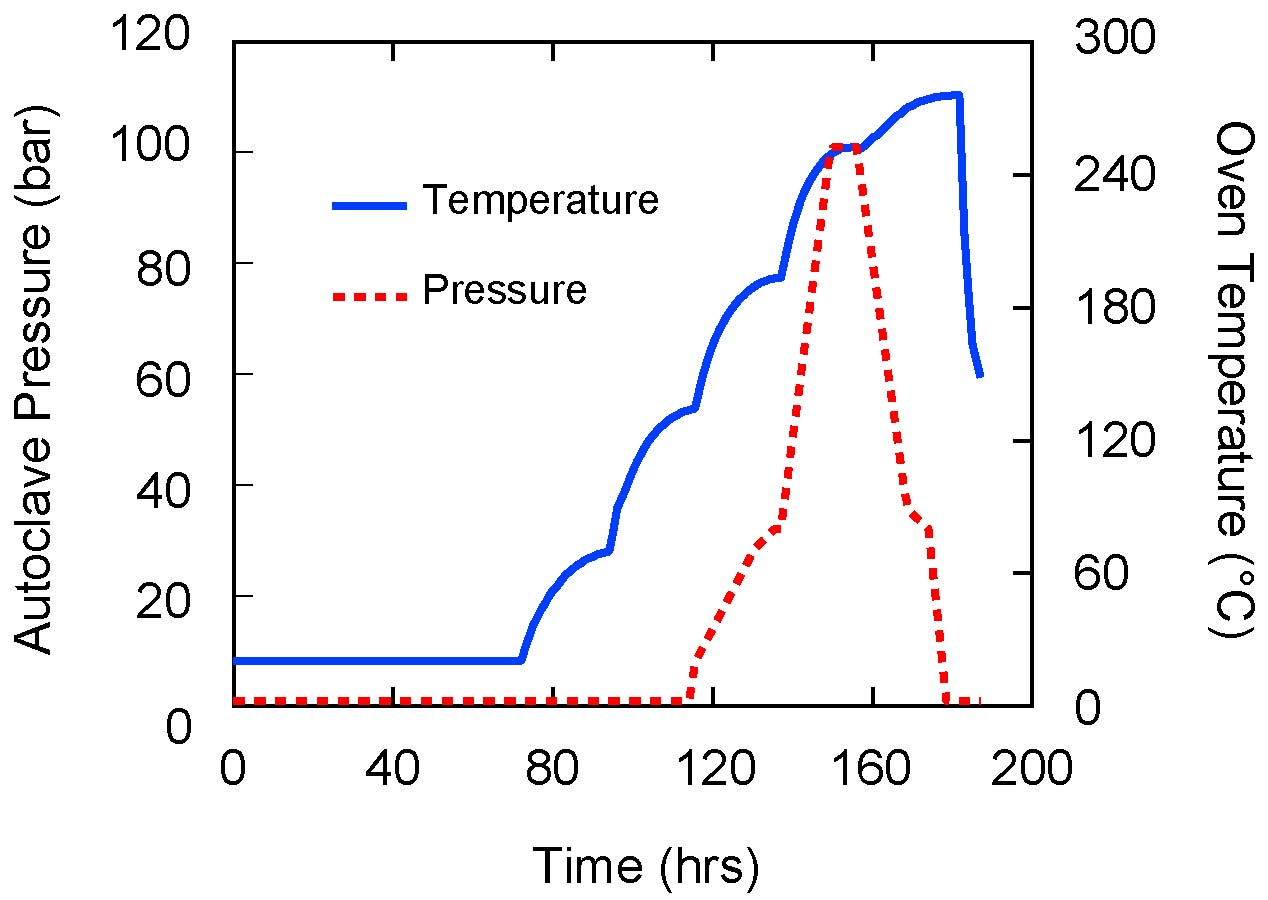}}
\caption{\label{fig1}(Color online) Oven temperature and autoclave pressure profile during aging and drying.}  
\end{figure}

Specifically, the sol is poured into a cylindrical stainless steel chamber having an inner diameter of 5.08 cm and a length of 4.92 cm and containing glass tubes of various sizes as molds.  The chamber is completely filled with the sol and sealed using two stainless steel lids each having a thickness of 1.27 cm.  Each lid is bolted in place using six 8-32 screws.  The alcogel that forms is typically aged at room temperature for approximately 3 days.  The chamber containing the alcogel is then loaded into an autoclave having an open volume of 0.46 L and containing $\sim 50$ mL of methanol.  The autoclave is loaded into an oven and heated through several manual adjustments and, as the temperature rises, the pressure in the alcogel chamber forces the lids to open slightly, allowing fluid to escape into the open volume of the autoclave.  Once the critical point of methanol ($T_c = 239.5^o$C, $p_c$ = 80.81 bar) is exceeded the autoclave is emptied by slowly opening a high pressure valve.  This drying typically begins from a starting temperature and pressure of $\sim 270^o$C and $\sim 100$ bar.  The system is depressurized $\sim 8-12$ hours and then
allowed to cool to room temperature overnight.  Fig.~\ref{fig1} depicts the typical oven temperature and autoclave pressure profiles during aging and drying.
With this method we have
successfully produced silica aerogel ranging in porosity from
$\sim 94\%-99\%$.  A sample's porosity is determined from its weight and volume after
supercritical drying to an accuracy of
0.1\% for a 98\%
aerogel.

For the SAXS and optical birefringence measurements reported here, cylindrical
samples with porosities ranging from $\sim 97\%-98\%$ were grown in glass tubes with $\sim 8$  mm inside diameter.  By controlling the
catalyst concentration
and the supercritical drying procedure we could grow samples with radial
shrinkage in the range $0\%$ to $20\%$.  From these samples we chose those with no significant shrinkage on-axis.  The samples were cut to $\sim 0.75$ cm with aspect ratios of $\sim 1$.

\section{SAXS EXPERIMENTS} 
The scattering experiments were
carried out at Sector 8 of the Advanced Photon Source (APS) at Argonne National
Laboratory, using a photon energy of 7.5 keV, corresponding to a wavelength,
$\lambda = 1.69$ {\AA}.  The X-ray beam, with a spot size of $\approx 100$ microns, was incident on the the center of a
sample, which was oriented with its cylinder axis perpendicular to the beam.  The experiments
discussed below can be grouped into two categories.  First, to study
the effect of
axial strain on the aerogel, scattering experiments were carried out on a nominally
isotropic sample as well as on a sample exhibiting 13.8\% radial shrinkage.  They
were subjected to increasing strain in the range 0\% to
$\sim 32\%$.  Secondly, in order to systematically explore the
relationship between radial
shrinkage and X-ray scattering, we conducted SAXS on a series of
samples exhibiting
radial shrinkage in the range $0\%$ to $20\%$.

\begin{figure}[h]
\centerline{\includegraphics[height=0.3\textheight]{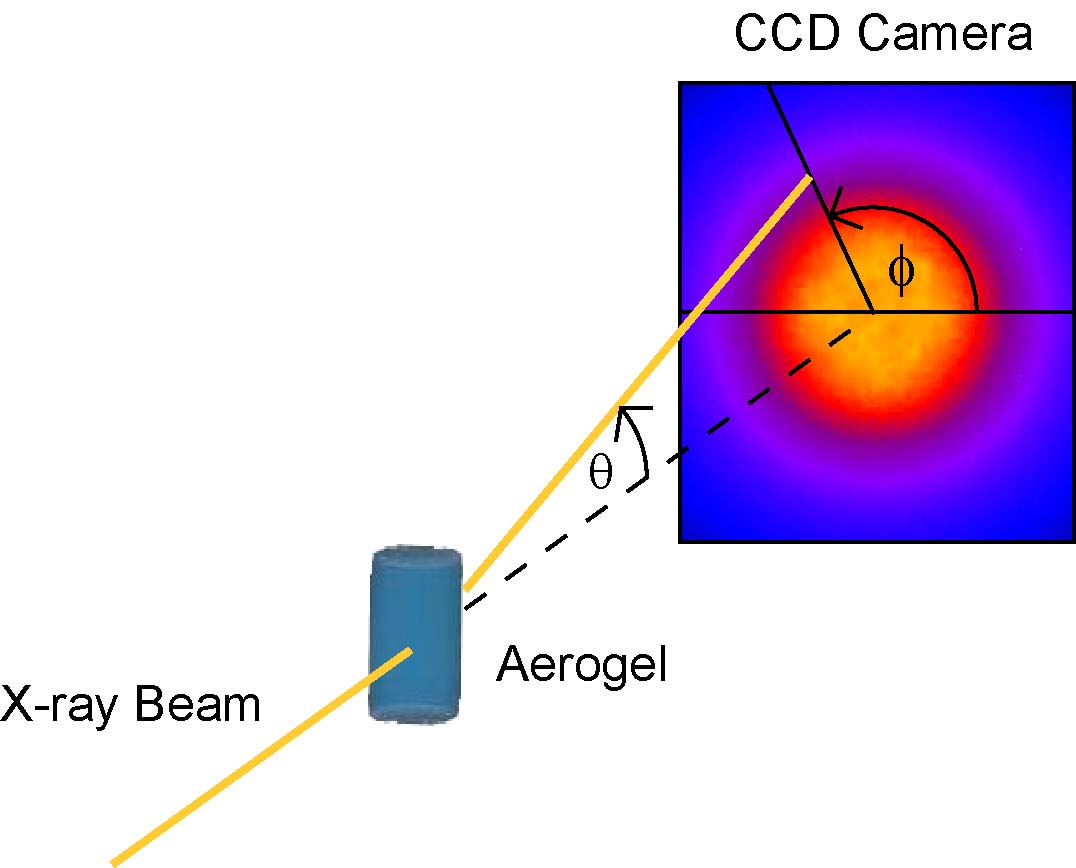}}
\caption{\label{fig2}(Color online) Sketch of the experimental
geometry. Note that $\phi =90^\circ$ is parallel to the cylinder axis.}
\end{figure}

In each experiment we recorded the scattered X-ray intensity, {\em
I}({\em q}), as a
function of the momentum transfer, {\em q}, using a CCD camera.  The momentum
transfer is defined as $q = (4\pi/\lambda) \sin(\theta/2)$, where
$\theta$ is the
angle between the incident and scattered X-ray wavevectors.  For a
given image, we
binned {\em I}({\em q}) in $\sim$10$^\circ$ increments of the azimuthal angle,
$\phi$, defined in the plane of the CCD camera (refer to
Fig.~\ref{fig2}).  The cylinder
axis was defined to be $\phi = 90^\circ$.  Fig.~\ref{fig2} is a
cartoon depicting the geometry of a typical SAXS measurement.

\section{SAXS RESULTS AND DISCUSSION I: Axial Strain and Anisotropy} 
\begin{figure}[h]
\centerline{\includegraphics[height=0.2\textheight]{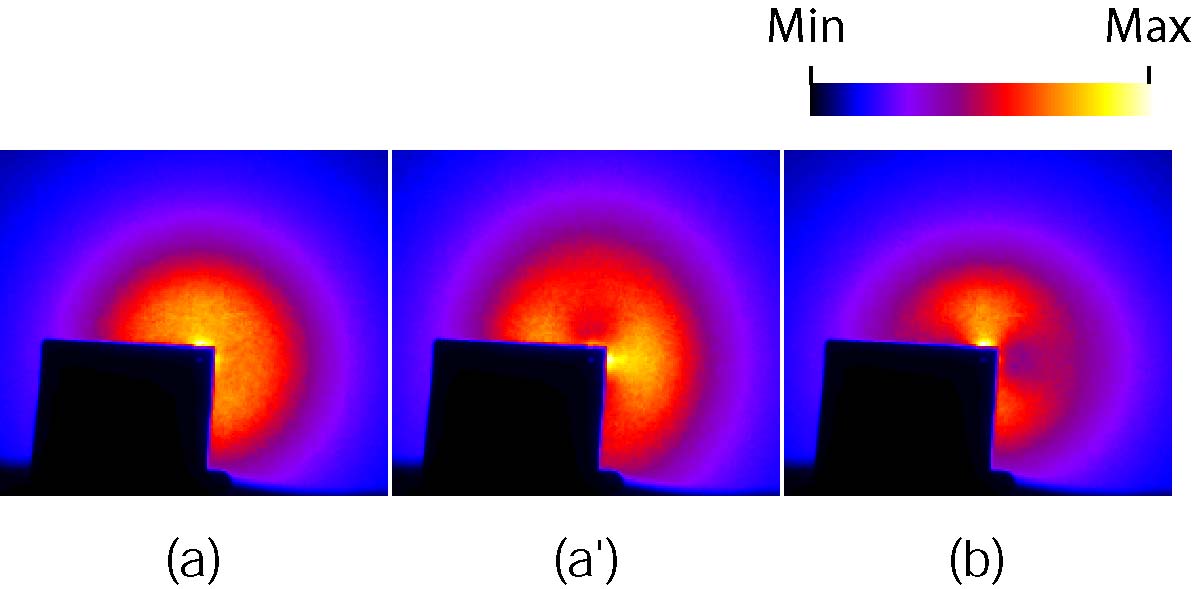}}
\caption{\label{fig3}(Color online) False color scattered X-ray intensities for (a) a $98\%$
nominally isotropic aerogel, (a$'$) the same sample unaxially strained by 12.7$\%$, and (b) a $97.6\%$ aerogel exhibiting 13.8\% radial shrinkage.
In each case the
sample is oriented with its cylinder axis vertical relative to the
images.  The dark
rectangle is a beamstop inserted to protect the CCD camera from the unscattered
transmitted beam.}
\end{figure}

The dependence of $I(q, \phi)$ on $\phi$ provides information
about the magnitude and direction of anisotropy in the aerogel.  This
is depicted in
Fig.~\ref{fig3}, which presents images of the scattered
X-ray intensity
from two aerogel samples.  The sample labeled (a) is of a nominally
isotropic and
unstrained aerogel, while (a$'$) depicts the same sample, but strained by
12.7\% along the cylinder axis.  The image labeled (b) is of a sample exhibiting 13.8\% radial
shrinkage.  In what follows, the labeling scheme presented in
Fig.~\ref{fig3} will
be used when referring to these two samples.  The porosity of sample
(a) was 98\%
and that of (b) 97.6\%.
Fig.~\ref{fig3} qualitatively demonstrates the existence of two
distinct types of anisotropy.  First, comparing (a) and (a$'$) shows
that axial
strain can be used to introduce anisotropy into a nominally isotropic
aerogel.  In
addition, (b) indicates the presence of intrinsic anisotropy in samples
exhibiting radial shrinkage.  Moreover, this intrinsic anisotropy is
$\sim 90^\circ$ out
of phase with that induced by axial strain.

\begin{figure}[h]
\centerline{\includegraphics[height=0.25\textheight]{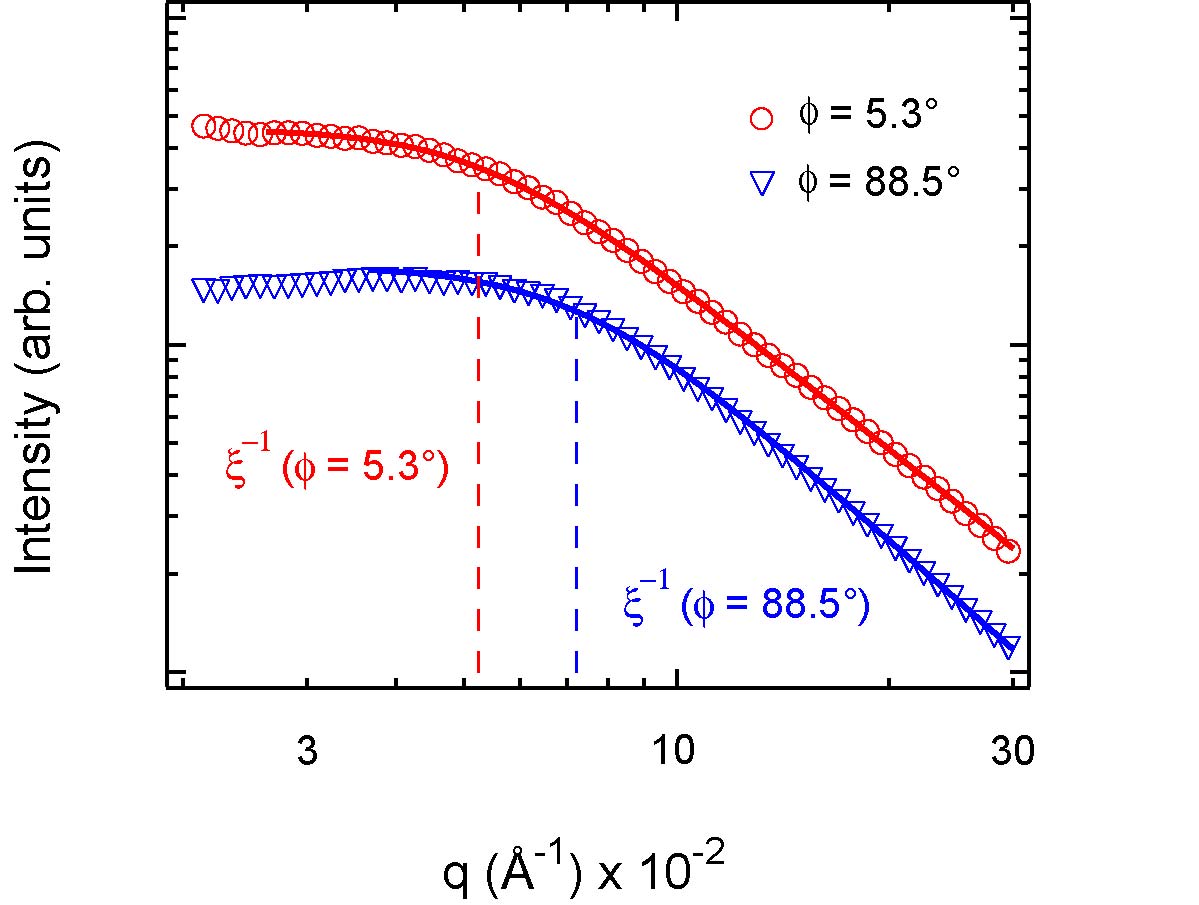}}
\caption{\label{fig4}(Color online) $I(q,\phi)$ fit with Eq. 1 ($\phi =
5.3^\circ$, $\phi = 88.5^\circ$) for the sample (a$'$).  The curves
have been offset
vertically for clarity, otherwise the data points would coincide at
high {\em q}.}
\end{figure}

We analyzed the scattering curves by fitting $I(q, \phi)$ with
the following phenomenological scattering function,
\begin{equation}
I(q,\phi)={{C\xi^{d}}\over{(1+q^{4}\xi^{4})^{d/4}}}{{(1+q^{2}\xi^{2})^{1/2}}\over{q\xi}}\sin[(d-1)\tan^{-1}(q\xi)],
\end{equation} where $C$, $d$, and $\xi$ are fit parameters which
depend on $\phi$.
Eq. 1 is a modified version of a structure factor for fractally
correlated materials
having an upper length scale cut-off.  The $q$-dependence of Eq. 1 is stronger than the structure factor derived by Freltoft {\it et. al.}\cite{Fre86}.  This sharper $q$-dependence was chosen to better match our experimental data.  The parameter $d$
is the fractal dimension of the aerogel, and $C$ is an overall scale
factor.  The length
scale
$\xi$ is associated with the aerogel correlation length, {\it i.e}. the
upper length scales
at which the aerogel ceases to be fractally correlated and, for future reference, we note that the axial (radial) correlation length corresponds to $\phi = 90^\circ$ ($\phi = 0^\circ$).    Note that in the
limit $q\xi \gg 1$, Eq. 1 is proportional to {$q^{-d}$} as
expected\cite{Fre86} for the fractal regime.  We performed fits to our data using
Eq. 1 and obtained an excellent representation as can be
seen in a typical case in Fig.~\ref{fig4} for sample (a$'$)
($\phi = 5.3^\circ$, $\phi = 88.5^\circ$).  Fig.~\ref{fig4} shows that when the aerogel is
compressed along the cylinder axis the
axial correlation length is
compressed relative to the correlation length in the radial direction.

\begin{figure}[t]
\centerline{\includegraphics[height=0.25\textheight]{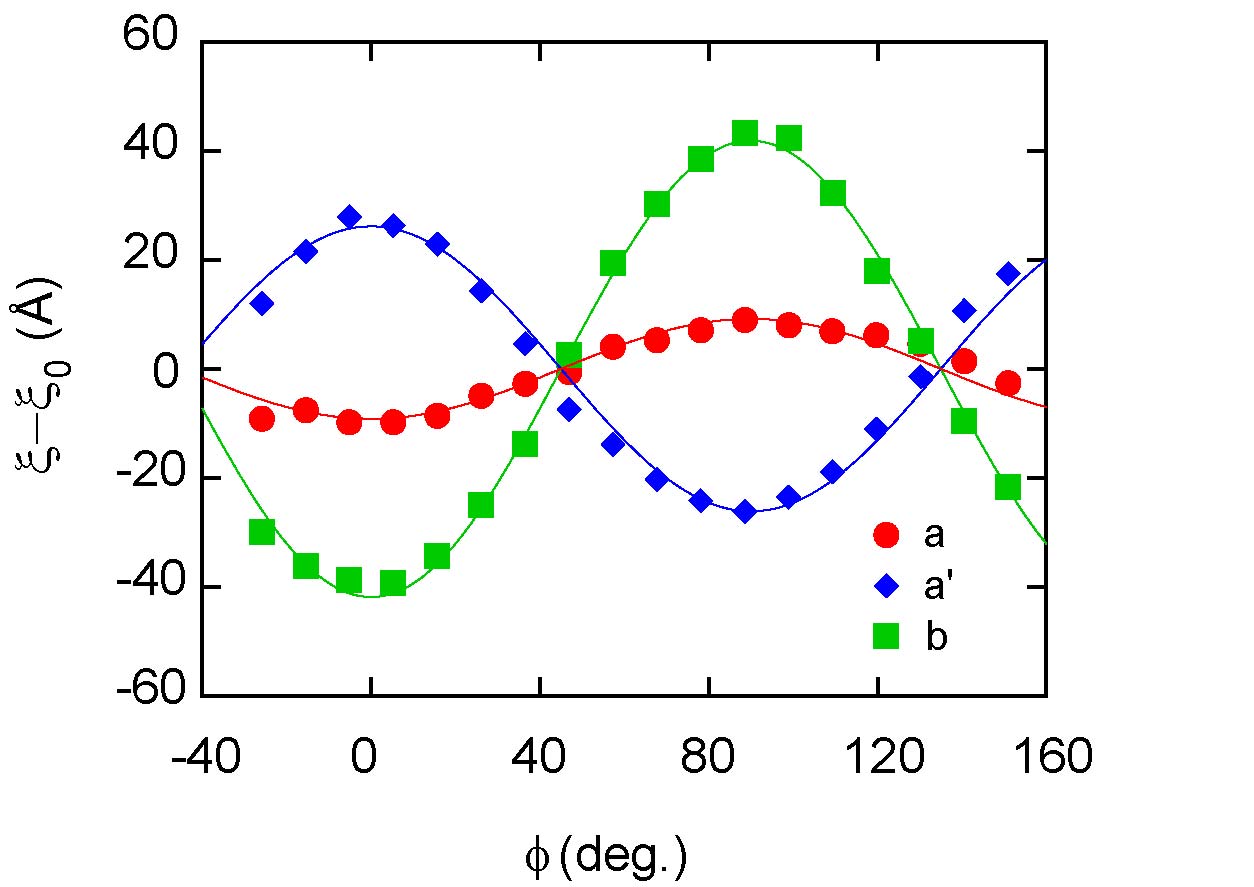}}
\caption{\label{fig5}(Color online) $\xi$-$\xi_{0}$ as a function of
$\phi$ for the
samples (a), (a$'$), and (b).  The solid curves are fits using Eq. 2.
The amplitude
of the sinusoid is a measure of the magnitude of anisotropy.}
\end{figure}

To quantify the magnitude and direction of the anisotropy in a given sample, we
plot the correlation length {$\xi$} versus {$\phi$}.  For all of the samples we
found the following $\phi$-dependence,
\begin{equation}
\xi(\phi)=\xi_{0}+\xi_{1}\cos(2\phi),
\end{equation} where $\xi_{0}$ and $\xi_{1}$ are fit parameters.  Note, the average fractal dimension ranged from 1.55-1.80 over the samples discussed in this work.  For any given sample, the fractal
dimension, $d$, was found to be a much weaker function of $\phi$, as compared to $\xi(\phi)$.  The
dependences  of $d$ and $C$ on
$\phi$ are not important to our analysis and do not pertain to our discussion.  The
parameter $\xi_{0}$ represents the average correlation length of the
aerogel and was
found to be of the order of $\sim$200 \AA.  The amplitude, $\xi_{1}$,
is a measure
of the magnitude of the anisotropy.  In Fig.~\ref{fig5} we present this
$\phi$-dependence for the samples (a), (a$'$), and (b).
The sinusoidal dependence of $\xi(\phi)$ demonstrates
that anisotropy from axial
strain, or radial shrinkage, exists on the length scale of the aerogel
correlation length, $\xi$.  Moreover, comparing the traces for
samples (a$'$) and
(b) demonstrates that the axial strain induced anisotropy has the opposite sign
of that from anisotropy produced by radial shrinkage.  Fig.~\ref{fig5}
also indicates
that the nominally isotropic sample, i.e. sample (a), also possesses
a small amount
of intrinsic anisotropy, which has a magnitude $\xi_{1}$ $\simeq$ 9
\AA.  

\begin{figure}[b]
\centerline{\includegraphics[height=0.25\textheight]{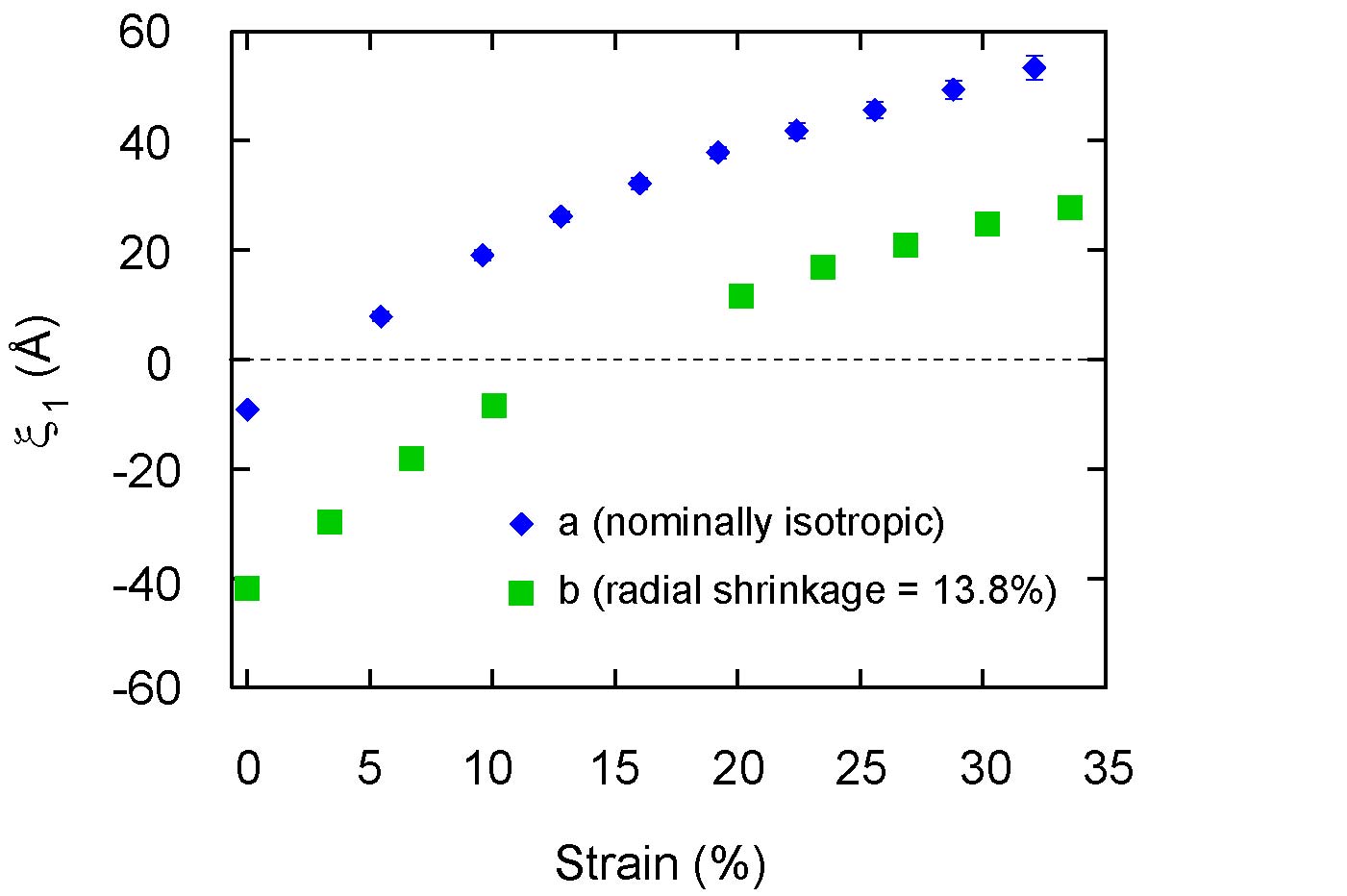}}
\caption{\label{fig6}(Color online) $\xi_{1}$ as a function of
axial strain applied to samples (a) and (b).  The vertical offset between the
two data sets reflects
the difference in magnitude of the intrinsic anisotropies of the two samples.}
\end{figure}

To further investigate the relationship between axial strain and global
anisotropy we plot $\xi_{1}$ versus applied axial strain.  The results for
samples (a) and (b) are presented in Fig.~\ref{fig6}.
\vspace{1cm}
The increase of $\xi_{1}$ with increasing strain demonstrates that
anisotropy can be introduced into the aerogel systematically.
Moreover, axial
strain can be used to tune between the two types of anisotropy and implies that
strain can be used to compensate for intrinsic anisotropy. The
negative values of $\xi_{1}$ are an indication of the overall phase in the
second term in Eq. 2.  

\begin{figure}[h]
\centerline{\includegraphics[height=0.25\textheight]{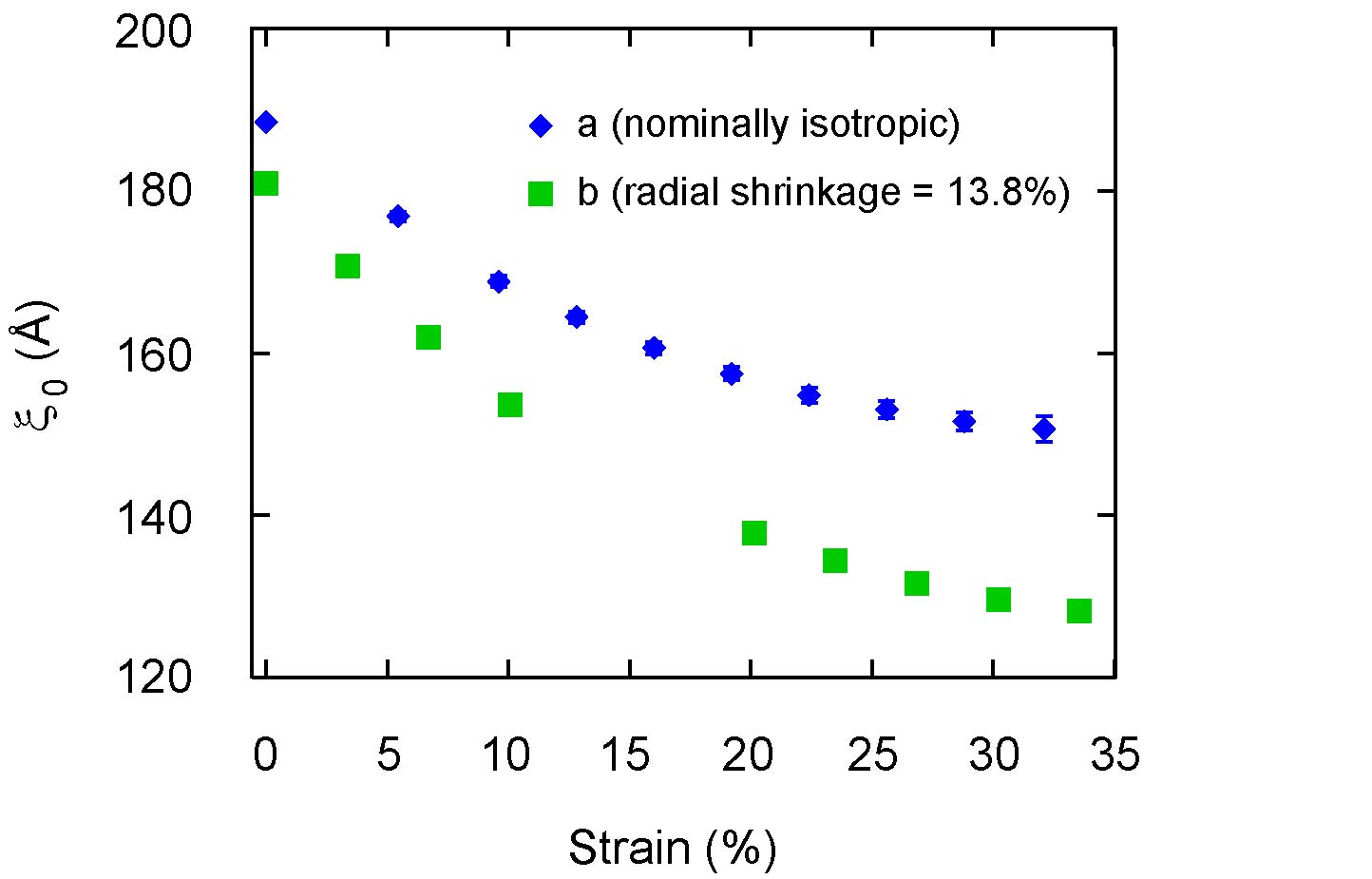}}
\caption{\label{fig7}(Color online) $\xi_{0}$ as a function of
axial strain applied to samples (a) and (b).}
\end{figure}

We also plot $\xi_{0}$ versus strain to see how the average
correlation length of a sample is affected.  The results
for samples (a)
and (b) are presented in Fig.~\ref{fig7}.  We find that $\xi_{0}$ decreases
monotonically with applied strain for both samples.  Previous SAXS studies on
isostatically strained aerogels also indicate a decrease in the
correlation length
with applied strain\cite{Woi99}.

\begin{figure}
\centerline{\includegraphics[height=0.45\textheight]{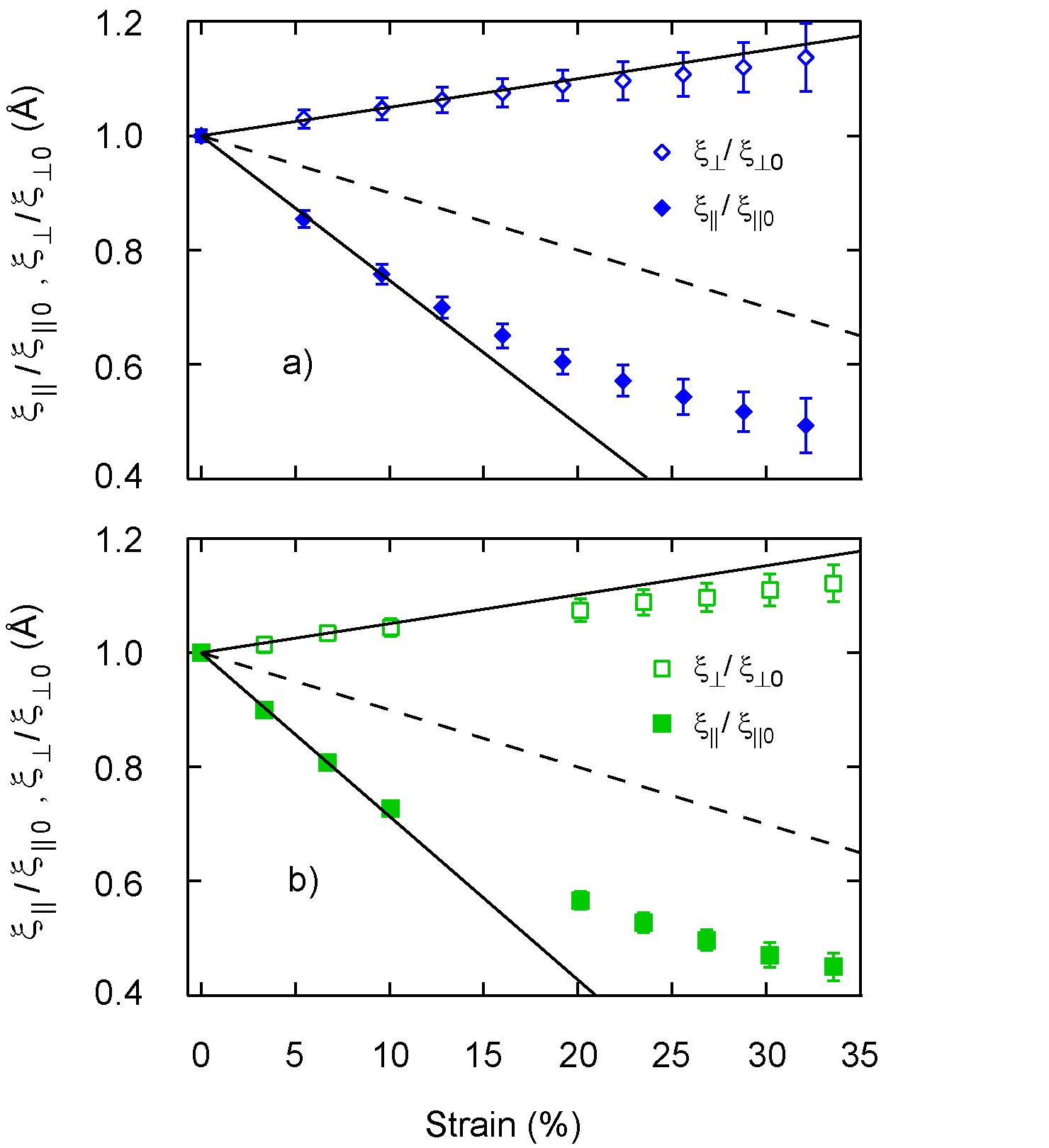}}
\caption{\label{fig8}(Color online) Normalized axial (radial) correlation length, $\xi_{\parallel}/\xi_{\parallel 0}$ ($\xi_{\perp}/\xi_{\perp 0}$), versus strain for sample (a) (upper panel) and (b) (lower panel).  The solid lines are linear fits to the low strain portion of $\xi_{\parallel}/\xi_{\parallel 0}$ and $\xi_{\perp}/\xi_{\perp 0}$.  The dashed line corresponds to $\alpha = 1$ in Eq. 3 and 4.}
\end{figure}

Fig.~\ref{fig6} and Fig.~\ref{fig7} seem to suggest that the primary effect of strain is to reduce the axial correlation length while only weakly increasing the radial correlation length.  The correlation length parallel (perpendicular) to the cylinder axis can be obtained from Eq. 2 by setting $\phi = 90^\circ$ ($\phi = 0^\circ$) and we define $\xi_{\parallel} \equiv \xi_0 - \xi_1$ ($\xi_{\perp} \equiv \xi_0 + \xi_1$) as the axial (radial) correlation length.  In Fig.~\ref{fig8} we plot $\xi_{\parallel}$ and $\xi_{\perp}$ divided by their values at zero strain, $\xi_{\parallel 0}$ and $\xi_{\perp 0}$, versus strain applied to samples (a) (uppper panel) and (b) (lower panel).
Fig.~\ref{fig8} clearly demonstrates that axial strain primarily compresses $\xi_{\parallel}$, {\it i.e.} the correlation length parallel to the strain axis, while only modestly increasing $\xi_{\perp}$ .

Fig.~\ref{fig8} also allows us to determine the extent to which the macroscopic strains are transmitted to microscopic length scales, {\it i.e.} length scales of the order of $\xi$.  For small values of strain the data in Fig.~\ref{fig8} indicate,
\begin{equation} 
\frac{\xi_{\parallel}}{\xi_{\parallel 0}} = (1+\alpha \epsilon_{\parallel})
\end{equation}
\begin{equation}
\frac{\xi_{\perp}}{\xi_{\perp 0}} = (1-\beta \epsilon_{\parallel}),
\end{equation}
where $\epsilon_{\parallel} \equiv \triangle l / l_{0}$ is the axial strain on a cylindrical aerogel intially having a length of $l_{0}$ and compressed by $\triangle l$.  The constants $\alpha$ and $\beta$ describe the transmission of macroscopic strain down to the level of the correlation length and are determined from linear fits of $\xi_{\parallel}/\xi_{\parallel 0}$ and $\xi_{\perp}/\xi_{\perp 0}$ near zero strain.  These fits are presented in Fig.~\ref{fig8} as solid lines for samples (a) (upper panel) and (b) (lower panel).  In Eq. 3 and 4 we use the convention that $\epsilon_{\parallel} < 0$ for compression.  Homogeneous elastic theory (HET) suggests that $\alpha = 1$ and $\beta = \nu$, where $\nu$ is the macroscopic Poisson ratio.  Note, $\alpha = 1$ implies that the macroscopic strain scales directly down to the level of the aerogel correlation length and is represented by the dashed lines in Fig.~\ref{fig8}.  If we had found $\alpha < 1$ it would suggest that strain is not distributed to microscopic length scales, while $\alpha > 1$ would imply an enhancement at microscopic length scales.  The dependence of $\xi_{\parallel}/\xi_{\parallel 0}$ on strain in Fig.~\ref{fig8} unambiguously indicates a value of $\alpha > 1$, implying that the aerogel cannot be accurately described by HET.  We find $\alpha = 2.5 \pm 0.1$ for sample (a) and $\alpha = 2.9 \pm 0.1$ for sample (b).  However, from our measurements we find that the ratio $\beta / \alpha \approx \nu$.  Specifically $\beta / \alpha = 0.20 \pm 0.01 $ for sample (a) and $ \beta / \alpha = 0.18 \pm 0.02$ for sample (b).  These values are qualitatively consistent with typical values of $\nu$ for silica aerogels\cite{Gro88}.  In addition, we have measured the macroscopic Poisson ratio for a sample similar to (a) using optical techniques (see section VII) and find $\nu = 0.30 \pm 0.05$.

We conclude that aerogel is not a homogenous elastic medium and that axial strain induces larger effects on the correlation length scale than would be expected for such a medium.  Finally, a Poisson-like behavior describing the microscopic displacements appears to hold.

\section{SAXS RESULTS AND DISCUSSION II: Intrinsic Anisotropy and Radial Shrinkage} 
\begin{figure}[h]
\centerline{\includegraphics[height=0.25\textheight]{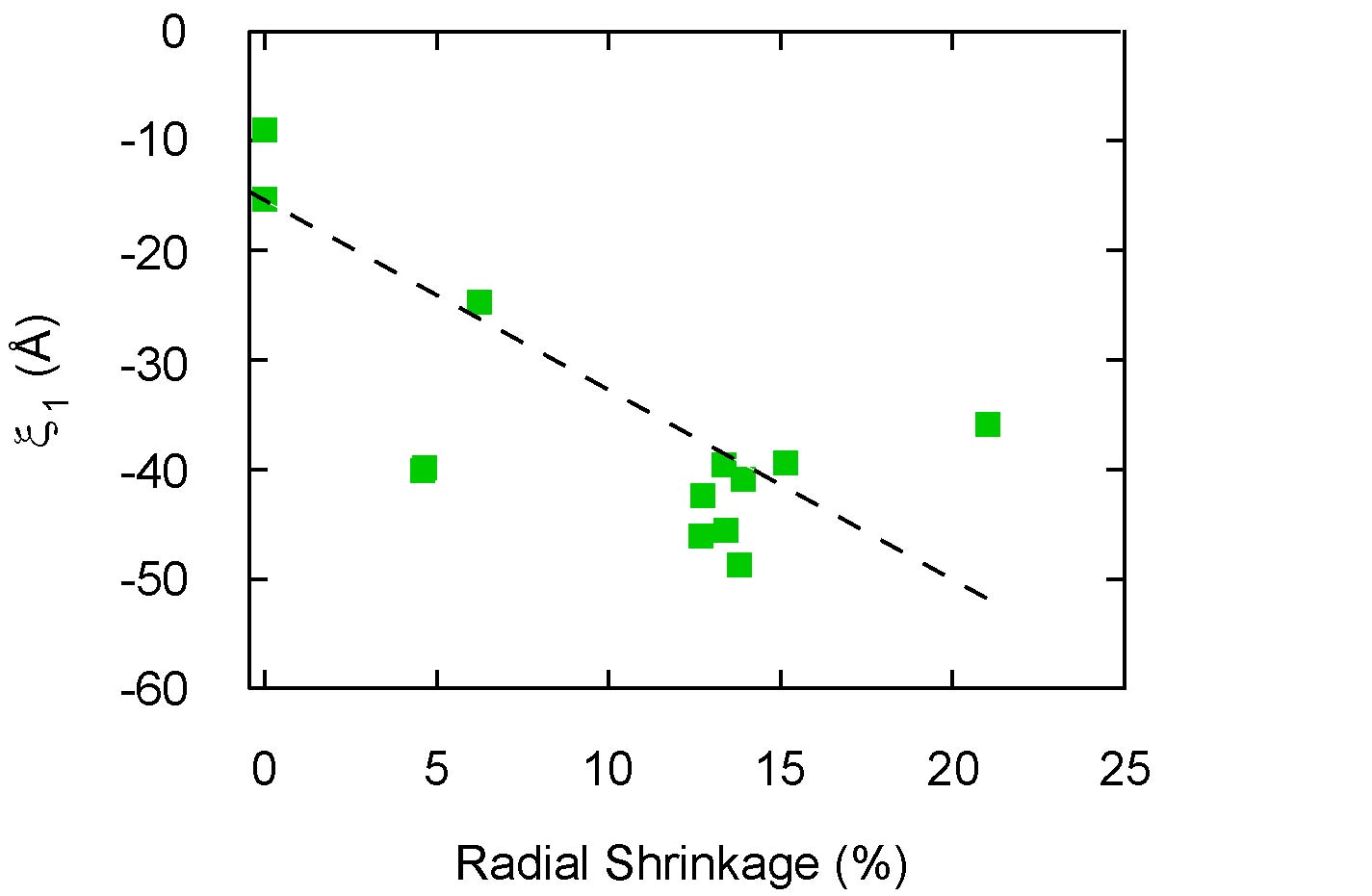}}
\caption{\label{fig9}(Color online) Anisotropy amplitude, $\xi_{1}$,
as a function
of radial shrinkage.  The dashed line is a guide to the eye. Note the relatively small magnitude of intrinsic
anisotropy
in the samples that did not exhibit shrinkage.}
\end{figure}

To explore the possibility of systematically introducing
anisotropy different from that produced by strain, we have also
performed SAXS
on a series of aerogels which exhibited preferential radial shrinkage after
supercritical drying.  Following the same analysis as detailed in the previous
section, we studied the evolution of the the anisotropy amplitude,
$\xi_{1}$, as a
function of radial shrinkage.  The results are presented in Fig.~\ref{fig9}.
Specifically, the absolute value of $\xi_{1}$ appears to increase
systematically with increasing shrinkage, indicating a
correlation between radial
shrinkage and anisotropy.  The larger data scatter in Fig. 7 as compared with Fig.
5, reflects that our measurement and control of radial shrinkage
is not as uniform and systematic as for imposed strain on a single sample.  However, the small anisotropy of the samples having small shrinkage is prominent.  It is noteworthy that the magnitude of the effect on $|\xi_{1}|$ is comparable with
that observed for strained samples.

\begin{figure}[h]
\centerline{\includegraphics[height=0.25\textheight]{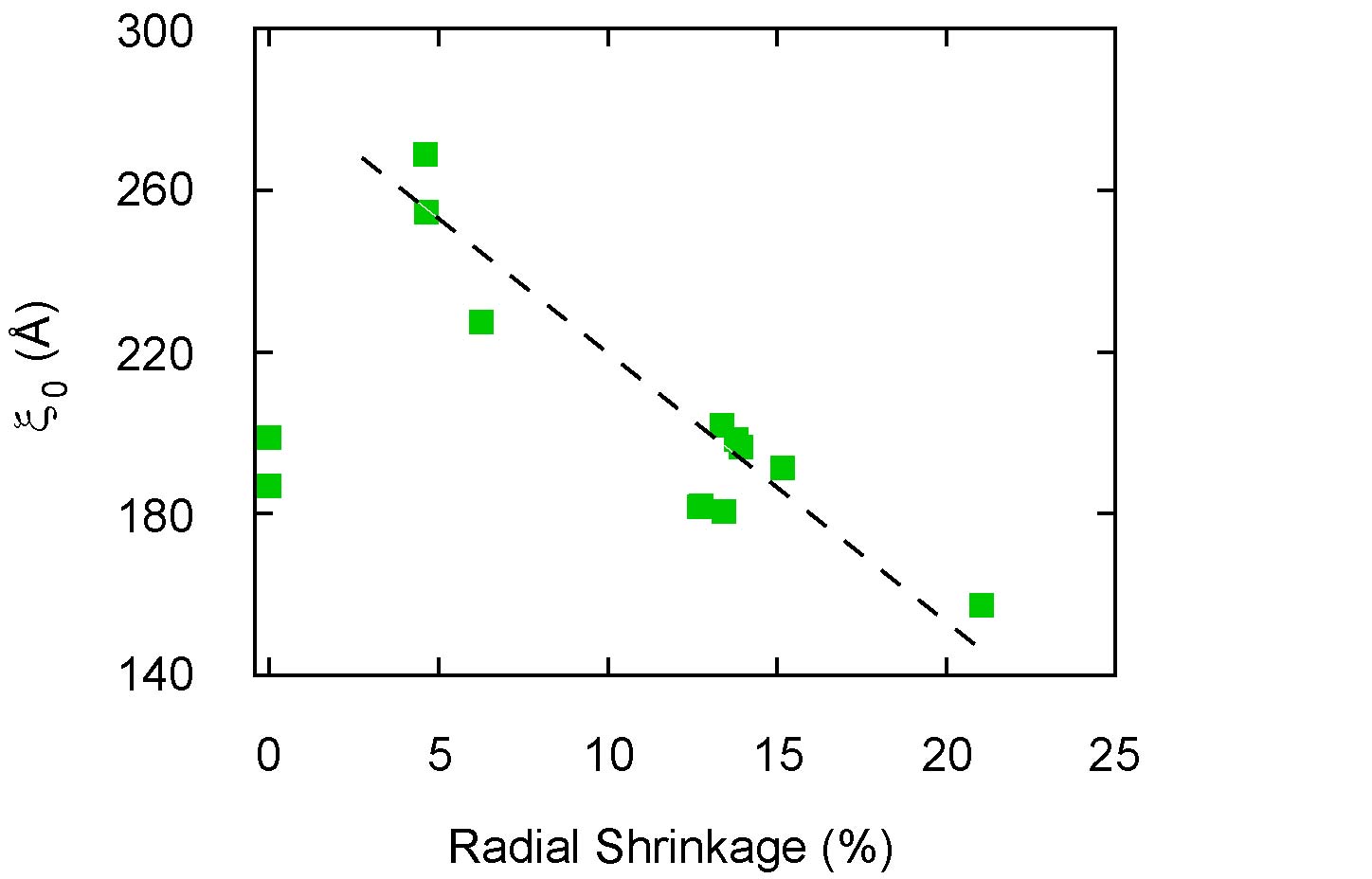}}
\caption{\label{fig10}(Color online) Average correlation length, $\xi_{0}$, as a
function of radial shrinkage.  The dashed line is a guide to the eye.  Note, the behavior of the samples exhibiting no shrinkage is inconsistent with the
general trend presented here, for which we have no explanation.}
\end{figure}

In addition, we plot the average correlation length,
$\xi_{0}$, as a function of radial shrinkage in Fig.~\ref{fig10}.
For the
samples that exhibited radial shrinkage, $\xi_{0}$ was found to decrease by as much as $\sim$40\%. This effect is of similar magnitude as that observed for strained samples
with the exception of the samples that did not show any shrinkage and which are not consistent with
the the general trend in  Fig.~\ref{fig10}, for reasons we do not understand.

\section{OPTICAL BIREFRINGENCE EXPERIMENTS} 
When a
transparent
material possesses an anisotropic
dielectric constant it will
exhibit birefringence.  Illuminating an optically birefringent material with white light will result in transmission of two components with orthogonal
polarizations, parallel and perpendicular to the ``optical axis''.
By placing such a material between crossed polarizers it is possible
to determine
if there is a well defined optical axis and how it is distributed
throughout the sample\cite{Hec98}.  Measurements of optical
birefringence are well-known
for characterization of liquid crystals\cite{Gra62} and many other
materials\cite{Bal70}.

We have performed optical birefringence experiments on high porosity aerogels.
Cylindrical aerogel samples were placed between two crossed polarizers and
illuminated with diffuse white light.  The cylinder axis was oriented vertically, a geometry similar to
that shown in Fig.~\ref{fig1}.  A diffuser was placed between the light source and the first
polarizer and images were
recorded with a digital camera located after the second polarizer.

To study the effect of axial strain on a nominally isotropic
aerogel, optical
birefringence experiments were conducted as the sample was
subjected to increasing strain in the range 0\% to 18.6\%.  In addition,
we have investigated samples with radial shrinkage and have
confirmed that optical
birefringence can be used to detect global anisotropy in this situation as well.
In both cases, the optical axis is primarily oriented along the cylinder axis.

\section{OPTICAL BIREFRINGENCE RESULTS AND DISCUSSION}
\begin{figure}[h]
\centerline{\includegraphics[height=0.1\textheight]{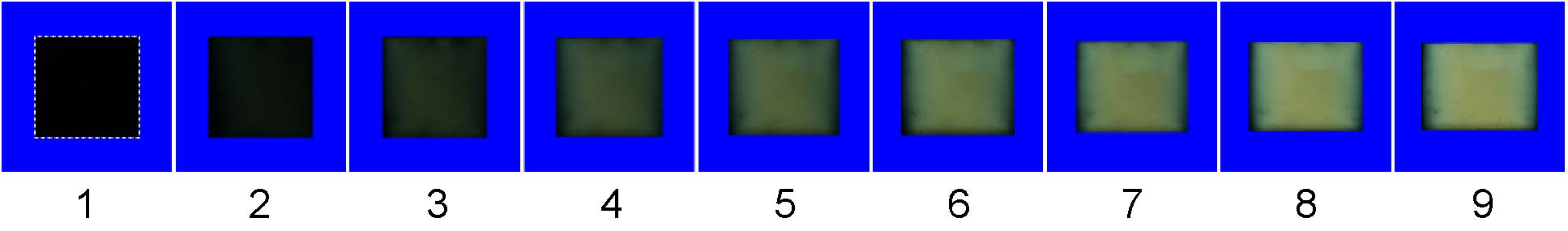}}
\caption{\label{fig11}(Color online) Optical birefringence of a $98\%$ porosity
aerogel, nominally isotropic before it was subjected to increasing axial strain.  The strain
increases from left to
right: 1 (unstrained), 2 (2.3\% strain), 3 (4.7\%), 4 (7.0\%), 5 (9.3\%), 6
(11.6\%), 7 (14.0\%), 8 (16.3\%), 9 (18.6\%).}
\end{figure}

\begin{figure}[h]
\centerline{\includegraphics[height=0.25\textheight]{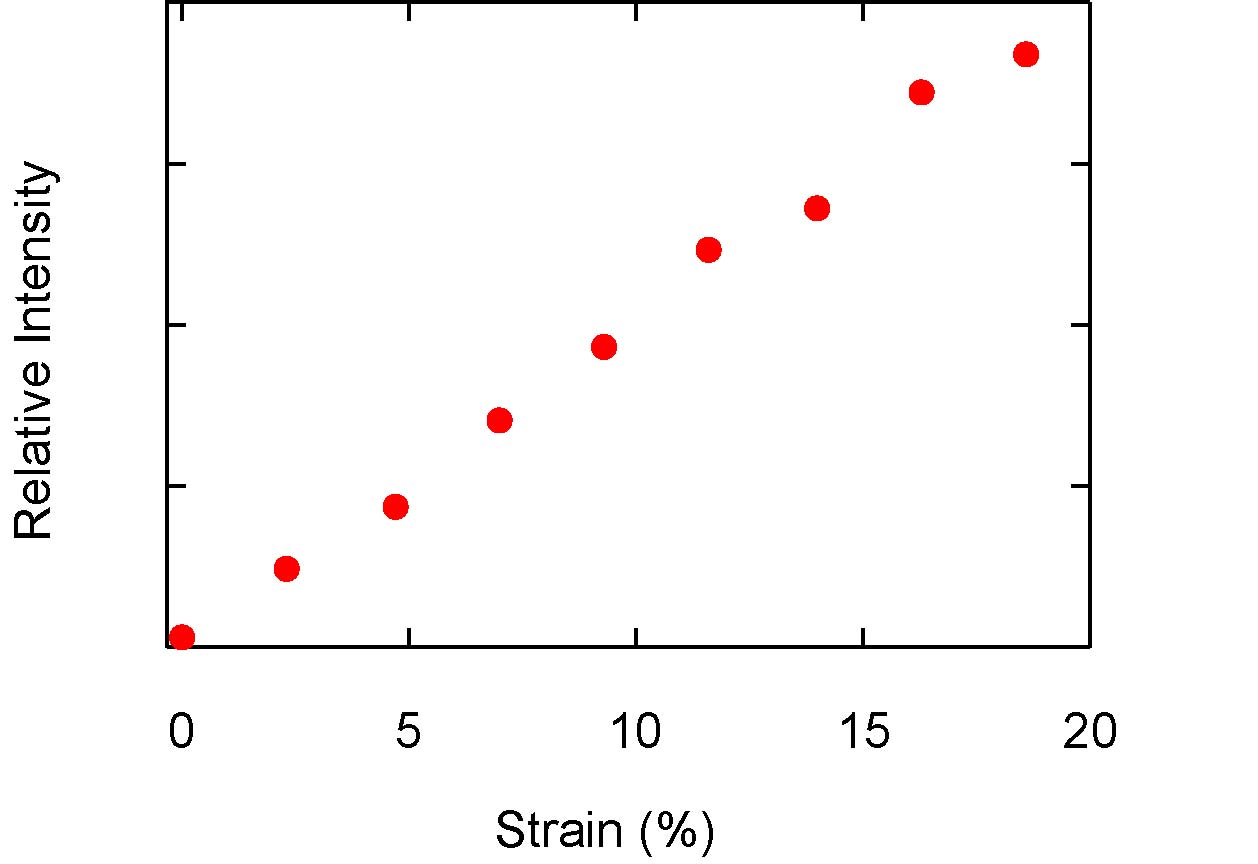}}
\caption{\label{fig12}(Color online) Relative intensity averaged over a central region of the image of the sample versus strain for the sample presented in Fig.~\ref{fig11}.}
\end{figure}

Fig.~\ref{fig11} presents optical birefringence of a
$98\%$ porosity aerogel as it was axially strained to
18.6\% in  2.3\% steps.  The crossed polarizers were oriented with one polarizer at 45$^\circ$ and
the other at 135$^\circ$ relative to the vertical cylinder axis.  This sample did not exhibit any radial shrinkage after supercritical drying.
Panel 1 of Fig.~\ref{fig11} indicates that the unstrained gel is homogeneously isotropic since it does not impose any preferred direction of polarization when placed between crossed polarizers for all orientations of the polarizers with respect to the cylinder axis.  We have also found no transmission of light propagating down the cylinder axis for this isotropic aerogel
placed between crossed polarizers.  Panels 2-9 of Fig.~\ref{fig11} demonstrate that
strain can be used to effectively convert this nominally isotropic aerogel into a polarizer.  We attribute this effect to optical birefringence.  The polarizing effect increases with increasing strain and is the result of
structural anisotropies on optical length scales.  Furthermore, it is
evident from the approximately uniform intensity of the image that this anisotropy is a global property of the entire sample and is homogeneously distributed.  From the images presented in Fig.~\ref{fig11} we can measure the relative intensity of light passing through the sample averaged over a central region of the image of the aerogel as a function of strain.  These results are presented in Fig.~\ref{fig12} and indicate that the transmitted intensity increases linearly with increasing strain up to the largest value of strain, {\it i.e.} $18.6\%$.
     
\begin{figure}[h]
\centerline{\includegraphics[height=0.1\textheight]{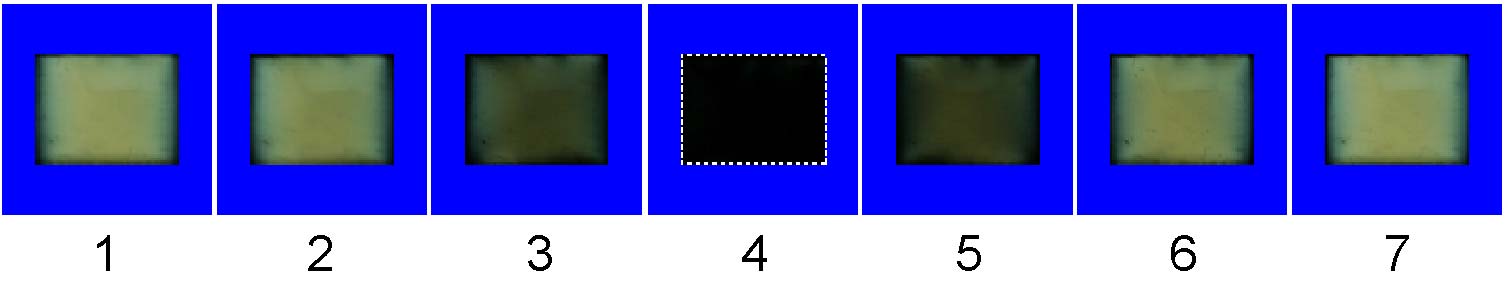}}
\caption{\label{fig13}(Color online) Optical birefringence of a $98\%$ porosity
aerogel strained by 18.6\%.  The labels are associated with the rotation of the
polarizers relative to the cylinder axis: 1 (45$^\circ$, 135$^\circ$), 2
(60$^\circ$, 150$^\circ$), 3 (75$^\circ$, 165$^\circ$), 4
(90$^\circ$, 180$^\circ$),
5 (105$^\circ$, 195$^\circ$), 6 (120$^\circ$, 210$^\circ$), 7 (135$^\circ$,
225$^\circ$).}
\end{figure}

By rotating both polarizers, keeping them crossed, we can determine the direction of the anisotropy axis.  Fig.~\ref{fig13}
presents such a rotation
sequence for the sample at it's maximum compression, i.e. 18.6\%.
The fact that the intensity maxima (minima) are seen when the
polarizers are oriented
at 45$^\circ$ and 135$^\circ$ (90$^\circ$ and 180$^\circ$) is consistent with the anisotropy axis being oriented along the cylinder (strain), axis.

\begin{figure}[h]
\centerline{\includegraphics[height=0.1\textheight]{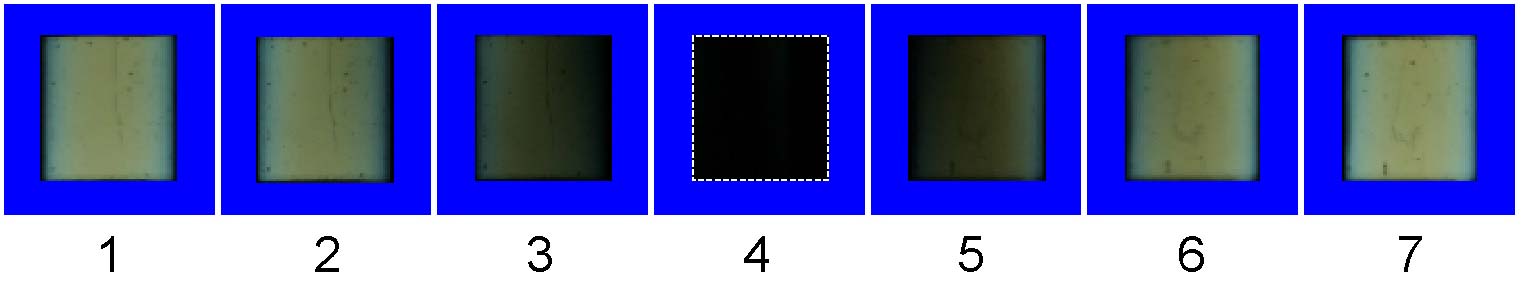}}
\caption{\label{fig14}(Color online) Optical birefringence of a $98\%$ porosity
aerogel exhibiting 12.7\% radial shrinkage.  The labels are
associated with the rotation
of the polarizers relative to the cylinder axis: 1 (45$^\circ$, 135$^\circ$), 2
(60$^\circ$, 150$^\circ$), 3 (75$^\circ$, 165$^\circ$), 4
(90$^\circ$, 180$^\circ$),5 (105$^\circ$, 195$^\circ$), 6 (120$^\circ$, 210$^\circ$), 7 (135$^\circ$,
225$^\circ$).}
\end{figure}

To complement our
SAXS measurements, we also used optical birefringence to investigate
possible intrinsic
anisotropy in a sample exhibiting radial shrinkage.  As before, the
polarizers were
intially oriented with one at 45$^\circ$ and the other at 135$^\circ$
relative to
the cylinder axis.  They were then rotated together, keeping them crossed.  We have found that aerogels exhibiting radial shrinkage polarize transmitted
light and this can be observed with cross polarizers, indicating the existence of a well defined optical axis.  Again we attribute this effect to
optical birefringence.
In Fig.~\ref{fig14} we present the results for an aerogel cylinder
exhibiting 12.7\%
radial shrinkage showing that axial intrinsic anisotropy is
present in this sample.
It is noteworthy that this intrinsic anisotropy seems to be less
uniform throughout the
sample than that induced by strain as displayed in panels 2, 3, 5, and 6 of Fig.~\ref{fig14}.

For the intrinsically anisotropic samples, {\em i.e.} those
exhibiting  radial shrinkage, we observed only a small amount of transmission through crossed polarizers for light propagating
down the cylinder axis.   This, along with the rotation series presented in Fig.~\ref{fig14}, indicates that the cylinder axis is predominantly the optical
axis in the case of samples exhibiting anisotropy due to radial shrinkage.

Our SAXS measurements indicate that microscopic anisotropy is associated with strain or radial
shrinkage developed during the sample growth and
drying stages. Furthermore, with a SAXS spot
size of $\approx 100$ microns it is possible to image the distribution of anisotropy by
scanning the sample profile.  We have found that tabletop optical polarization studies are
complementary and less time consuming.  They give a clear picture of the principal directions of the anisotropy on
optical length scales and have the advantage of displaying an image that shows the homogeneity of
this anisotropy as it is distributed over the sample.

\section{CONCLUSIONS} 
In conclusion, our
measurements demonstrate
that high porosity silica aerogels can be engineered with global structural anisotropy produced either
by applied  strain or during
sample growth and drying.  SAXS
measurements indicate that this anisotropy is on the length scale of
the aerogel
correlation length,
$\xi$, and optical birefringence measurements confirm that it is a global feature of the aerogel.  These aerogels represent a novel type of anisotropic porous
medium with
possible implications for applications in a wide variety of physical systems.

\begin{acknowledgments} 
We would like to thank J.A. Sauls for
valuable theoretical
insights.  We are grateful to N. Mulders, G.W. Scherer, and J.F. Poco for their
advice regarding aerogel shrinkage and fabrication.  We are indebted
to P.E. Wolf and
F. Bonnet for introducing us to the technique of optical polarization studies.  B. Reddy
and W.J. Gannon have provided many useful discussions.

This work was
supported by the National Science Foundation, DMR-0703656.  Use of
the Advanced Photon
Source was supported by the U. S. Department of Energy, Office of
Science, Office of
Basic Energy Sciences, under Contract No. W-31-109-ENG-38.
\end{acknowledgments}

\end{document}